# Detection and imaging of atmospheric radio flashes from cosmic ray air showers

H. Falcke[4,9,1], W. D. Apel[6], A. F. Badea[6], L. Bähren[4], K. Bekk[6], A. Bercuci[3], M. Bertaina[11], P. L. Biermann[1], J. Blümer[6,5], H. Bozdog[6], I. M. Brancus[3], S. Buitink[9], M. Brüggemann[10], P. Buchholz[10], H. Butcher[4], A. Chiavassa[11], K. Daumiller[6], A. G. de Bruyn[4], C. M. de Vos[4], F. Di Pierro[11], P. Doll[6], R. Engel[6], H. Gemmeke[7], P. L. Ghia[12], R. Glasstetter[13], C. Grupen[10], A. Haungs[6], D. Heck[6], J. R. Hörandel[5], A. Horneffer[1], T. Huege[1], K.-H. Kampert[13], G. W. Kant[4], U. Klein[2], Y. Kolotaev[10], Y. Koopman[3], O. Krömer[7], J. Kuijpers[9], S. Lafebre[9], G. Maier[6]*, H. J. Mathes[6], H. J. Mayer[6], J. Milke[6], B. Mitrica[3], C. Morello[12], G. Navarra[11], S. Nehls[6], A. Nigl[9], R. Obenland[6], J. Oehlschläger[6], S. Ostapchenko[6], S. Over[10], H. J. Pepping[4], M. Petcu[3], J. Petrovic[9], S. Plewnia[6], H. Rebel[6], A. Risse[8], M. Roth[5], H. Schieler[6], G. Schoonderbeek[4], O. Sima[3], M. Stümpert[5], G. Toma[3], G. C. Trinchero[12], H. Ulrich[6], S. Valchierotti[11], J. van Buren[6], W. van Cappellen[4], W. Walkowiak[10], A. Weindl[6], S. Wijnholds[4], J. Wochele[6], J. Zabierowski[8], J. A. Zensus[1] & D. Zimmermann[10]

[1]*Max-Planck-Institut für Radioastronomie, 53121 Bonn, Germany.* [2]*Radioastronomisches Institut der Universität Bonn, 53121 Bonn, Germany.* [3]*National Institute of Physics and Nuclear Engineering,7690 Bucharest, Romania.* [4]*ASTRON, 7990 AA Dwingeloo, The Netherlands.* [5]*Institut für Experimentelle Kernphysik, Universität Karlsruhe, 76021 Karlsruhe, Germany.* [6]*Institut für Kernphysik, Forschungszentrum Karlsruhe, 76021 Karlsruhe, Germany.* [7]*Institut für Prozessdatenverarbeitung und Elektronik, FZK, 76021 Karlsruhe, Germany.* [8]*Soltan Institute for Nuclear Studies, 90950 Lodz, Poland.* [9]*Department of Astrophysics, Radboud University, 6525 ED Nijmegen, The Netherlands.* [10]*Fachbereich Physik, Universität Siegen, 57072 Siegen, Germany.* [11]*Dipartimento di Fisica Generale dell'Università, 10125 Torino, Italy.* [12]*Istituto di Fisica dello Spazio Interplanetario, INAF, 10133 Torino, Italy.* [13]*Fachbereich C - Physik, Universität Wuppertal, 42097 Wuppertal, Germany.* *Present address: University of Leeds, Leeds LS2 9JT, UK.*

**The nature of ultrahigh-energy cosmic rays (UHECRs) at energies >$10^{20}$ eV remains a mystery[1]. They are likely to be of extragalactic origin, but should be absorbed within ~50 Mpc through interactions with the cosmic microwave background. As there are no sufficient powerful accelerators within this distance from the Galaxy, explanations for UHECRs range from unusual astrophysical sources to exotic string physics[2]. Also unclear is whether UHECRs consist of protons, heavy nuclei, neutrinos or γ-rays. To resolve these questions, larger detectors with higher duty cycles and which combine multiple detection techniques[3] are needed. Radio emission from UHECRs, on the other hand, is unaffected by attenuation, has a high duty cycle, gives calorimetric measurements and provides high directional accuracy. Here we report the detection of radio**

**flashes from cosmic-ray air showers using low-cost digital radio receivers. We show that the radiation can be understood in terms of the geosynchrotron effect[4–8]. Our results show that it should be possible to determine the nature and composition of UHECRs with combined radio and particle detectors, and to detect the ultrahigh-energy neutrinos expected from flavour mixing[9,10].**

When UHECRs interact with particles in the Earth's atmosphere, they produce a shower of elementary particles propagating towards the ground with almost the speed of light. The first suggestion that these air showers also could produce radio emission was made[11] on the basis of a charge-excess mechanism, which is very strong for showers developing in solid media[12,13]. In a couple of experimental activities in the 1960s and 1970s, coincidences between radio pulses and cosmic ray events were indeed reported[14,15]. Owing to the limitations of electronics in those days, the measurements were cumbersome and did not lead to useful relations between radio emission and air shower parameters. As a consequence, the method was not pursued for a long time and the historic results came into question. However, the mechanism for the radio emission of air showers was recently revisited and proposed to be coherent geosynchrotron emission[4]: Secondary electrons and positrons produced in the particle cascade rush with velocities close to the speed of light through the Earth's magnetic field and are deflected. As in synchrotron radiation, this produces dipole radiation that is relativistically beamed into the forward direction. The shower front emitting the radiation has a thickness that is comparable to (or less than) a wavelength for radio emission below ~100 MHz. Hence the emission is expected to be coherent to a large extent, which greatly amplifies the signal.

To see whether radio emission from cosmic rays is indeed detectable and useful in a modern cosmic ray experiment, we have built the LOPES (LOFAR Prototype Station) experiment. LOPES is a phased array of dipole antennas with digital electronics developed to test aspects of the LOFAR (Low-Frequency Array) concept. Compared to historical experiments, it provides an order of magnitude increase in bandwidth and time resolution, effective digital filtering methods, and for the first time true interferometric imaging capabilities. The radio array is co-located with the KASCADE[16] (Karlsruhe Shower Core and Array Detector) experiment that is now part of KASCADE-Grande at the research centre in Karlsruhe, Germany (see Supplementary Fig. 1). KASCADE provides coincidence triggers for LOPES and well-calibrated information about air shower properties. Experimental procedure and data reduction have been described elsewhere[17] and we give here only a brief summary in the Methods section. A related experiment is currently under way at the Nançay radio observatory[18].

Using LOPES, we have detected the radio emission from cosmic ray air showers at 43–73 MHz on a regular basis with unsurpassed spatial and temporal resolution (see Supplementary Fig. 2). After digital filtering of radio interference we still have almost the full bandwidth of $\Delta\nu$=33 MHz available, giving us a time resolution of $\Delta t \approx 1/\Delta\nu$=30 ns compared to ~1 µs resolution in historic experiments. Using radio interferometric techniques we can also image the radio flash for the first time. An example is shown in Fig. 1, where the air shower is the brightest radio point source on the sky for some tens of nanoseconds (see also Supplementary Video 1). The nominal resolution in our maps is ~2° in azimuth and elevation towards the zenith. Within these limits the emission appears point-like. To put this in perspective, we note that previous radio experiments used fixed analogue beams with a width of ~20° and no possibilities for imaging. Current detectors for UHECRs (for example, AUGER) are limited to about ~1° accuracy. Positional accuracies for LOPES can be a fraction of a degree for bright sources. This will improve further with interferometer baselines longer than used here.

To make an initial and reliable statistical assessment of the radio properties of air showers, we have investigated a rather restrictive set of events with relatively high signal-to-noise ratio and simple selection criteria. The criteria are purely based on shower parameters reconstructed from KASCADE. Using events from the first half year of operation, starting January 2004, we selected all events with a shower core within 70 m of the centre of LOPES, a zenith angle <45°, and a reconstructed 'truncated muon number' of $N_\mu$>$10^{5.6}$≈4×$10^5$. The truncated muon number is the reconstructed number of muons within 40–200 m of the shower core. For KASCADE, this quantity is a good tracer of primary particle energy[19], $E_p \propto N_\mu^{0.9}$. The selection corresponds approximately to $E_p$>$10^{17}$ eV. This is the upper end of the energies that KASCADE was built for. Using these selection criteria ('cuts') leaves us with 15 events and a 100% detection efficiency of the radio signal. This avoids any bias due to non-detections. The rather restrictive cut on the shower core location allows us to ignore radial dependencies. Also, the antenna gain reduces significantly at zenith angles >45°. Even though neither KASCADE nor LOPES are optimized for large zenith angles, we have also checked for highly inclined events. Selecting all events with a much lower truncated muon number of $N_\mu$>$10^5$ and zenith angle >50° we still detect >50% of all events — in many cases with very high field strengths. However, given the current uncertainties of KASCADE in shower parameters for inclined showers, we ignore those events for our analysis below.

The strength of the detected radio pulses in our sample is some µV per m per MHz at present, but we still lack an accurate absolute gain calibration. The position of the radio flashes are coincident with the direction of the shower axis derived from KASCADE data within the errors. The average offset is (0.8±0.4)° as determined from our radio maps.

We find the strongest correlations between the absolute value of the electric field strength height of the pulse, $\varepsilon$, and $N_\mu$, and between $\varepsilon$ and the geomagnetic angle, $\alpha_B$. The latter is defined as the angle between the shower axis and the geomagnetic field. No obvious correlation is found with the zenith angle. Theory[4] predicts that owing to coherence the electric field strength should scale linearly with the number of particles in the shower, which — for KASCADE — is approximately proportional to the number of muons. Hence, to first order we can separate the two effects by dividing the electric field of the radio pulse by the truncated muon number. We find that $\varepsilon_\mu = \varepsilon/N_\mu \propto (1-\cos\alpha_B)$, where the lowest geomagnetic angle in the sample was $\alpha_B=8°$ (see Supplementary Fig. 3). It is also possible to use a $\varepsilon_\mu \propto \sin\alpha_B$ behaviour, which for our data gives only marginally worse results. Simulations[6,7] indicate that this strong dependence on $\alpha_B$ is largely a polarization effect, as LOPES antennas currently measure only a single polarization in the east–west direction.

We can use this dependence to correct for the geomagnetic angle. Figure 2a shows the measured radio signal $\varepsilon$ plotted against muon number, while Fig. 2b shows the same plot where we use a normalized field strength height, $\varepsilon_\alpha$, corrected for the $(1-\cos\alpha_B)$ dependence. The correlation, albeit with only few events, clearly improves and shows a remarkably low scatter of ±16%. For an error-weighted nonlinear fit we find that $\varepsilon_\alpha \propto N_\mu^{1.2\pm0.1}$, which, within the errors, is consistent with a linear relation, $\varepsilon_\alpha \propto E_p$.

The close association between radio flashes and cosmic ray air showers in direction and time shows that the radio emission is directly associated with the shower itself. Our selection of events based purely on shower parameters has yielded a 100% detection efficiency, suggesting that radio emission from extensive air showers is a common and reliable tracer of cosmic rays. As predicted, the electric field of the radio emission is coherent. The good correlation with the geomagnetic angle demonstrates that the emission is caused by the interaction of the shower with the Earth's magnetic field. All results that we have found so far match the basic predictions for the geosynchrotron effect.

The most encouraging result is the very tight correlation of the radio emission with the primary particle energy. The essentially linear increase of the electric field strength with energy is a consequence of coherence, and hence the received radio power will increase quadratically with primary particle energy, making this technique particularly suitable for cosmic rays at much higher energies than studied here. For covering large detector areas—to increase count rates at the high-energy end of the UHECR spectrum—radio antennas are particularly suitable, owing to their ease of deployment and low cost.

Radio detection of cosmic rays with digital phased arrays has a number of features that make it superior (or complementary) to current techniques for UHECR detection in a number of areas. For example, the good directional accuracy reachable with radio interferometers will significantly improve clustering studies in search for UHECR point sources if used in a large ground array like LOFAR.

Another very promising area involves composition studies. The number of electrons integrated over the entire shower, revealing the primary particle energy, is difficult to determine with particle detectors: only a small fraction of these electrons reach the ground, due to their short absorption length. Muons, on the other hand, reach the ground largely unharmed, and their number is higher for showers produced by an iron nucleus compared to proton-induced, or even γ-induced, showers. Hence, a combined radio- and muon-detector array could determine the spectrum and composition of UHECRs. Compared to hybrid detector arrays, where optical fluorescence techniques with ~10% duty cycle are used to obtain calorimetric measurements, the duty cycle for radio hybrid studies would be almost ten times higher. Hence radio arrays could be the way to study UHECR composition at the very highest energies, where much larger event rates are needed than are currently available.

A particularly intriguing application concerns highly inclined showers, which were expected to be very easily detected with radio antennas[5,6,20]. The predicted detectability is supported by our high detection rate of showers at large zenith angles, despite a lower threshold on muon number. Highly inclined showers travel through several times the vertical air mass, and very few shower particles reach a detector on the ground. The exceptions are neutrinos, which can travel large distances without interactions and generate inclined showers at any distance to the ground[9]. Showers induced by electron neutrinos will have a relatively low hadronic component relative to the leptonic component. Therefore, for inclined showers the ratio between the radio and the muon signal on the ground will be a tell-tale signal of electron neutrinos.

Additionally, it has been suggested that Earth-skimming tau neutrinos, produced through flavour mixing, will also lead to highly inclined air showers when the secondary tau decays after some 50 km travel length at $10^{18}$ eV (ref. 10). No hadronic or photon showers are expected at such low inclinations. Hence, radio antennas with sensitivity towards the horizon would also be ideal tau neutrino detectors and provide clues about flavour mixing. Although the first detection of ultrahigh-energy neutrino events would be extremely exciting in itself, the calorimetric and far-looking nature of radio detection would even allow a relatively reliable energy determination of electron and tau neutrino showers—something that had been considered extremely difficult in the past with surface detector arrays[10].

**Supplementary Information** accompanies the paper on www.nature.com/nature.

**Acknowledgements** A.F.B. is on leave of absence from the National Institute of Physics and Nuclear Engineering, Bucharest; T.H. is now at the Institut für Kernphysik, Forschungszentrum Karlsruhe, Karlsruhe; S.Os. is on leave of absence from Moscow State University, Moscow. LOPES was supported by a grant from the German Federal Ministry of Education and Research, under grant No. 05 CS1ERA/1 (Verbundforschung Astroteilchenphysik). This work is part of the research programme (grant 03PR2180) of the `Stichting voor Fundamenteel Onderzoek der Materie (FOM)', which is financially supported by the `Nederlandse Organisatie voor Wetenschappelijk Onderzoek (NWO)'. The KASCADE Experiment is supported by the German Federal Ministry of Education and Research. The Polish group is supported by KBN grant 1 P03B 03926(2004-06); The Romanian group acknowledge grants from the Romanian Ministry of Education and Research (CERES97/2004, PN0320/0205).

**Author Information** Reprints and permissions information is available at npg.nature.com/reprints&permissions. The authors declare no competing financial interests. Correspondence and requests for materials should be addressed to H.F. (falcke@astron.nl).

## Methods

**Description of the array and standard data processing**

The current set-up of LOPES consists of 10 'inverted-V' antennas distributed over the KASCADE array[16], which consists of 252 detector stations on a uniform grid with 13 m spacing, and which is electronically organized in 16 independent clusters. Each antenna is centred between four KASCADE huts in the northwestern part of the array. The antenna set-up has a maximum baseline of 125 m.

The analogue radio signal received by the antenna is filtered to select a band from 43 to 76 MHz, giving a bandwidth of $\Delta\nu$=33 MHz, and is digitized with a 12-bit, 80-MHz analogue-to-digital converter (ADC), that is, using the second Nyquist zone (40–80 MHz) of the ADC. The data are continuously written into a cyclic ring buffer with a buffering time of 6.7 s (1 Gbyte). About 0.8 ms of data are read out whenever KASCADE produces a 'large-event trigger', which means that 10 out of 16 KASCADE clusters have produced an internal trigger. The resulting trigger rate is about two per minute, giving a total data volume of about 3.5 Gbyte d$^{-1}$. The dead-time during read-out is ~0.6 s. On average the trigger is delayed with respect to the shower by 1.8 μs, depending on the shower geometry relative to the array.

We relate the radio signal to shower properties using parameters provided by the standard KASCADE data processing from the particle detectors[16]. These parameters are: location of the shower core, shower direction, energy deposited in the particle detectors, total number of electrons at ground level, and the reconstructed truncated muon number.

The basic processing of the data consists of several steps: correction of instrumental delays for each antenna using a TV transmitter with known position, digital filtering of narrow-band interference, frequency and elevation-dependent gain correction for each antenna, flagging of antennas with unusually high noise, and correction of trigger delays based on shower direction. Delay corrections are done by applying a phase gradient to the Fourier transform of the time series data. Digital filtering is achieved by setting the amplitudes of narrow-band transmitters in the Fourier domain to an average value of the surrounding channels[17]. In the next step, a digital beam is formed in the direction of the shower axis by correcting for the delay of the light-travel time, and summing up the digitized and calibrated E-field time series data from each antenna ('adding beam'). Alternatively, one can also add up data from each unique combination of antennas where the E-field time series data of the antennas in each pair have been multiplied ('cross-correlation beam') before summing. To obtain the received power, the adding beam needs to be squared. Conversion back to 'historically' used units[15] of μV m$^{-1}$ MHz$^{-1}$—an integrated absolute pulse field strength height $\varepsilon$—is done by multiplying with $1/\Delta\nu$ and taking the square root of the data.

We fit a gaussian profile to the largest peak in a narrow time window around −1.8 μs from the trigger signal where the shower is expected to arrive. Finally, we check for the curvature of the wave front by correcting antenna delays assuming a spherical wave front with curvature radius $R_{cur}$, and iterating $R_{cur}$ until the radio peak is maximal.

The average full-width at half-maximum of the radio pulses in time was 49±10 ns. This is mostly due to broadening by the finite filter bandwidth, which we have not yet attempted to deconvolve. The average curvature radius of the wave front was $R_{cur}$=5,000±3,000 m, the latter being a statistical scatter. The curvature is rather poorly determined, as our longest baselines are an order of magnitude smaller than the average $R_{cur}$. Curvature has only a significant effect for events with high signal-to-noise ratio. For weak events, the change in amplitude for different $R_{cur}$ is less than the noise in the pulse height. An analysis using a fixed value of $R_{cur}$=2,500 m for each event was found to change the final result in this Letter only marginally.

In addition to the standard data reduction described above, we produced for each of the selected events a sky-map of the radio emission. The mapping routine calculates the radio intensity by adding the electric fields of the dipoles for a four-dimensional data grid as a function of azimuth, elevation, distance and time. We used a spatial resolution of 0.2° on the plane of the sky, distance slices separated by 250 m, and a time resolution of 12.5 ns. Within this four-dimensional data cube we located the intensity maximum to obtain the radio properties of the shower, and fitted a gaussian to the time profile. The resolution of the interferometer beam (~2°) becomes asymmetric and degrades in elevation for sources towards the horizon, as we have a planar array.

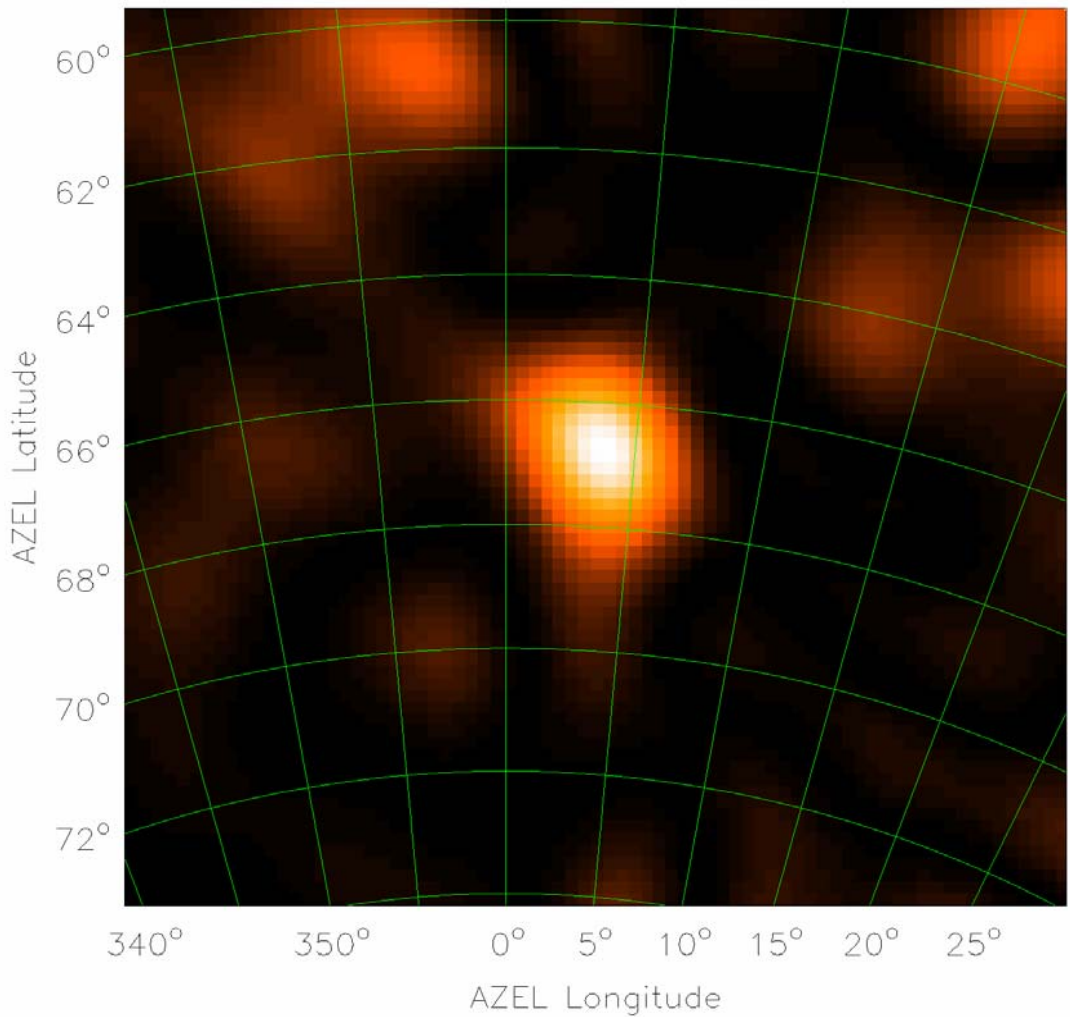

**Figure 1 | Radio map of an air shower.** For each pixel in the map, we formed a beam in this direction integrated over 12.5 ns and show the resulting electric field intensity. Longitude and latitude give the azimuth and elevation (AZEL) direction (north is to the top, east to the right). The map is focused towards a distance of 2,000 m (fixed curvature radius for each pixel). The cosmic ray event is seen as a bright blob of 2.4°×1.8° size. Most of the noise in this map is due to interferometer sidelobes caused by the sparse radio array. No image deconvolution has been performed.

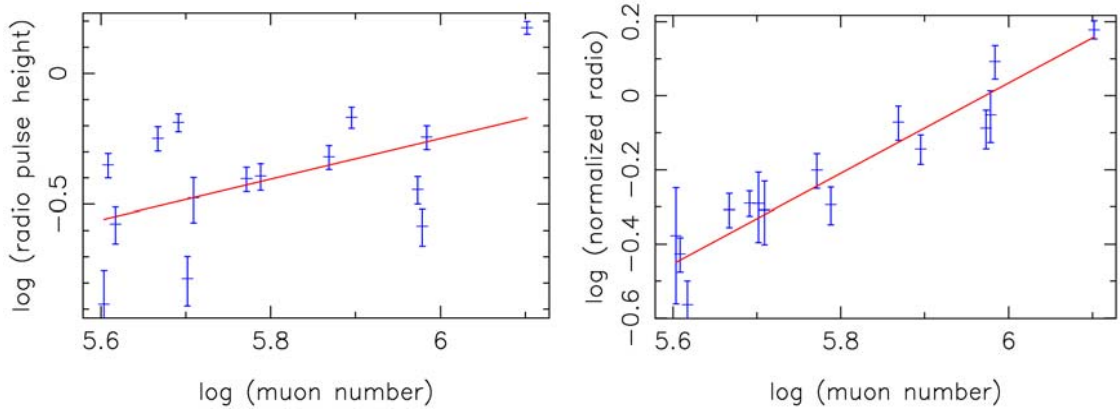

**Figure 2 | Radio emission as a function of muon number. a**, The logarithm of the radio pulse height, $\varepsilon$, versus the logarithm of muon number, which has not been corrected for the geomagnetic angle dependence. **b**, As **a** but now the geomagnetic angle dependence is corrected for ($\varepsilon_\alpha$). The correlation improves significantly. Solid lines indicate power-law fits. Errors were calculated from the noise in the time series before the pulse plus a nominal 5% error on gain stability. Both errors were added in quadrature. The radio pulse height units are arbitrary.

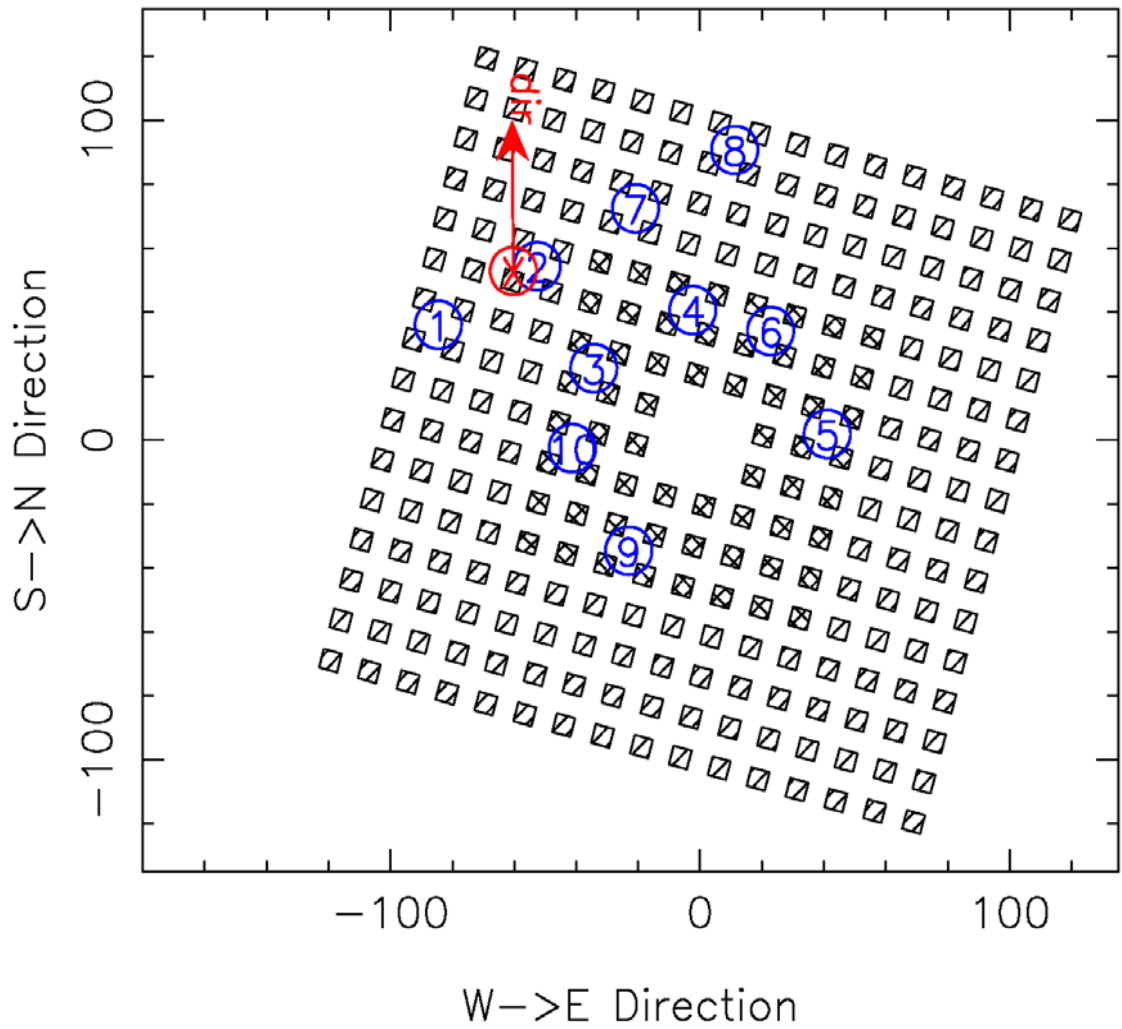

Supplementary Figure 1: Layout of the LOPES array. The axes are in meters; north is to the top. The boxes mark KASCADE detector huts. Double-hashed boxes indicate detectors with additional muon detectors. Blue circles show the location of the LOPES antennas. The red arrow shows the direction of the cosmic ray shower used as an example in this paper which has a zenith angle of 23.4°. The shower core impact location is identified by a red cross.

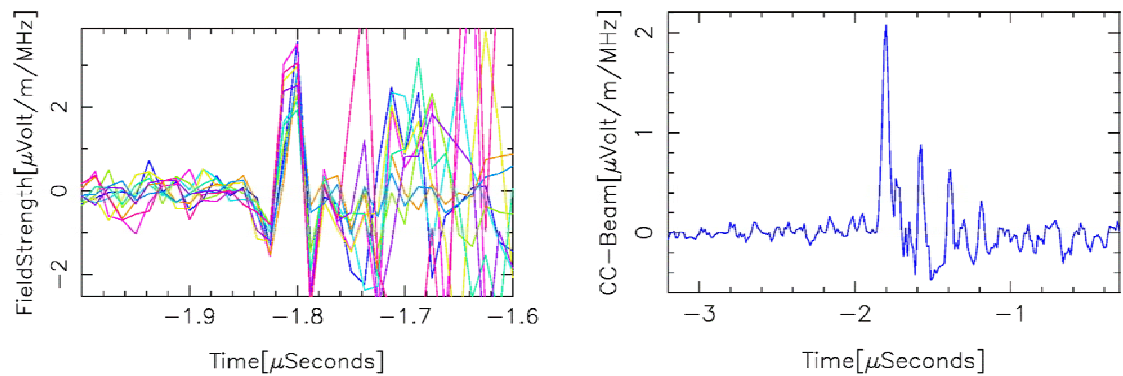

Supplementary Figure 2: Electric field as a function of time for our strongest event. The left panel (a) shows the electric field for each dipole after correcting for instrumental and geometric delays towards the air shower. The field is almost fully coherent at $t = -1.8\mu s$, the arrival time of the shower. The incoherent noise thereafter is local radio emission from the photo multipliers which is delayed due to the finite transport time within the dynode chain. The right panel (b) shows the radio emission as a function of time after beam-forming, i.e. multiplying all two-antenna combinations in (a) and summing the result. Since we are slightly over-sampling the available bandwidth and in order to remove the fine-structure due to the sampling frequency, we smooth (block-average) the beam-formed data with a time constant of $\Delta t$=37.5 ns, i.e. three samples, which is roughly ~ $1/\Delta \nu$. The incoherent photo multiplier radio noise is greatly reduced and the CR event stands out.

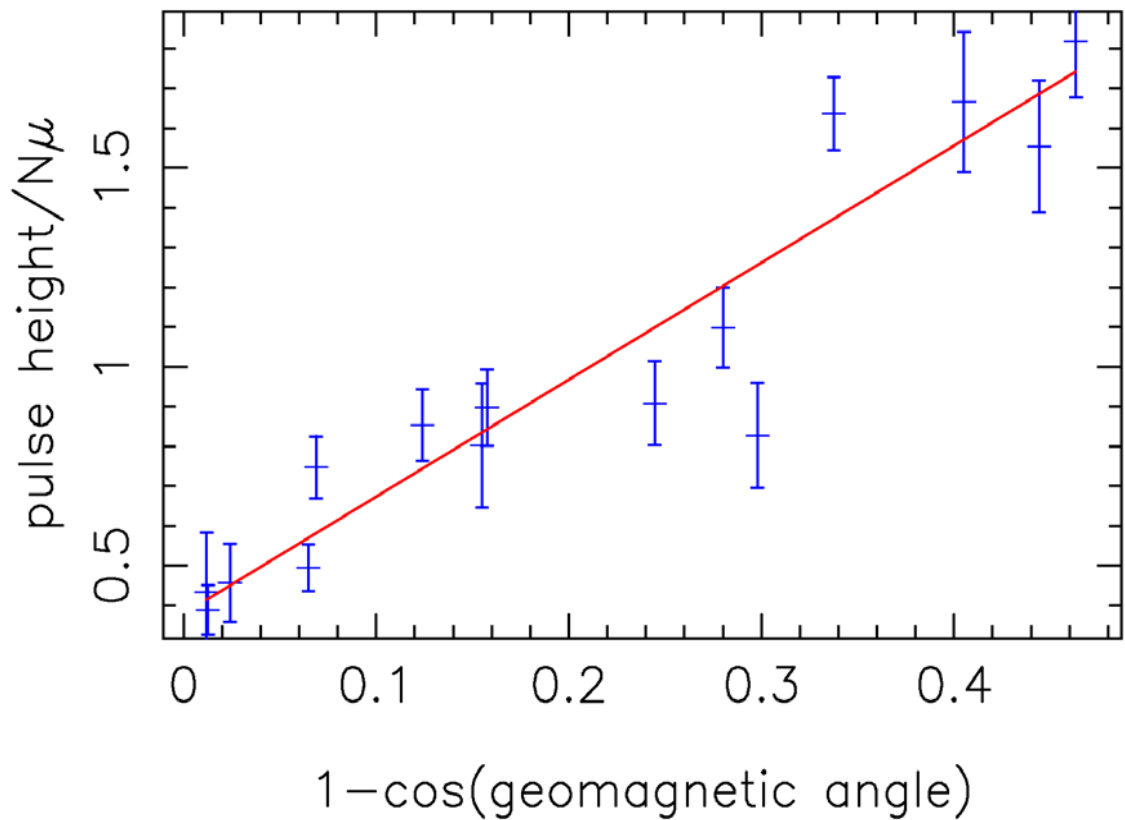

Supplementary Figure 3: Dependence of radio signal on geomagnetic angle. The panel shows the radio pulse height ε divided by the total number of muons in the shower plotted versus 1-cos($α_B$), where $α_B$ is the angle between the shower axis and the geomagnetic field.

*See separate movie file*

Supplementary Video 1: This movie shows an animated version of Figure 1 where the sky map is shown for different time steps before and after the event. The movies span 112.5 ns. Between consecutive data frames we have interpolated five frames for slowing it down. The data was Hanning smoothed in the time domain over three data frames.